\documentclass{article}
\usepackage[utf8]{inputenc}
\usepackage{geometry}
\usepackage{amsmath}
\usepackage{amssymb}
\usepackage{graphicx}
\usepackage{amsmath}
\usepackage{hyperref}
\usepackage{booktabs}
\usepackage{multirow} 
\usepackage{tabularx}
\usepackage{booktabs}
\usepackage{makecell} 
\usepackage{geometry} 
\usepackage{caption}
\usepackage{soul}
\usepackage{subcaption}
\usepackage{hyperref}

\usepackage{booktabs}
\usepackage[numbers, sort&compress]{natbib}
\title{PiRD: Physics-informed Residual Diffusion \\for Flow Field Reconstruction}
\author{
Siming Shan$^a$, Pengkai Wang$^a$, Song Chen$^b$, Jiaxu Liu$^b$, Chao Xu$^a$, Shengze Cai$^a$\thanks{Corresponding author: shengze\_cai@zju.edu.cn.} \\
~\\
$^a$ Institute of Cyber-Systems \& Control, 
College of Control Science \& Engineering, \\
Zhejiang University, Hangzhou, 310027, China  \\
$^b$ School of Mathematical Sciences, Zhejiang University, Hangzhou, 310027, China  \\
}
\date{}
\begin{document}

\maketitle

\begin{abstract}
The use of machine learning in fluid dynamics is becoming more common to expedite the computation when solving forward and inverse problems of partial differential equations. 
Yet, a notable challenge with existing convolutional neural network (CNN)-based methods for data fidelity enhancement is their reliance on specific low-fidelity data patterns and distributions during the training phase. In addition, the CNN-based method essentially treats the flow reconstruction task as a computer vision task that prioritizes the element-wise precision which lacks a physical and mathematical explanation. 
This dependence can dramatically affect the models' effectiveness in real-world scenarios, especially when the low-fidelity input deviates from the training data or contains noise not accounted for during training. The introduction of diffusion models in this context shows promise for improving performance and generalizability. Unlike direct mapping from a specific low-fidelity to a high-fidelity distribution, diffusion models learn to transition from any low-fidelity distribution towards a high-fidelity one. Our proposed model - Physics-informed Residual Diffusion, demonstrates the capability to elevate the quality of data from both standard low-fidelity inputs, to low-fidelity inputs with injected Gaussian noise, and randomly collected samples. By integrating physics-based insights into the objective function, it further refines the accuracy and the fidelity of the inferred high-quality data. Experimental results have shown that our approach can effectively reconstruct high-quality outcomes for two-dimensional turbulent flows from a range of low-fidelity input conditions without requiring retraining. 
\end{abstract}

\section{Introduction}
As a typical inverse problem, reconstructing spatial fields from sparse sensor data is a highly challenging task, yet its applications are widespread, encompassing numerous fields such as medicine, industrial devices, remote sensing, fluid dynamics, and more \cite{course2023state, erichson2020shallow, shen2015missing, ravishankar2019image, davoudi2019deep, zhao2022sparse}.
Efforts to improve traditional methodologies for effectively handling sparse input data have to face a balance between computational efficiency and effectiveness \cite{brunton2020machine}.
Meanwhile, due to the potential movement and loss of sparse sensors, the fidelity of experimental setups utilizing these collected data is inevitably affected, which leads to structural instability \cite{wu2023fast}.
Moreover, the noise introduced by the sensor measurement process cannot be ignored. 
All these aspects lead to a complex challenge for reconstruction, which is difficult to meet by traditional methods.

In fluid dynamics, Direct Numerical Simulation (DNS) \cite{lee2015direct} is a classic method that simulates fluid motion directly by solving the complete Navier-Stokes equations in all scales.
To obtain high-fidelity flow field information, DNS requires significant computational resources, especially when solving ill-posed problems or inverse problems related to turbulent flows.
Indeed, owing to computational efficiency challenges, several methods employing simplified turbulence models for approximate calculations have been devised, such as Large Eddy Simulation (LES) \cite{piomelli1999large}, Reynolds-Averaged Navier-Stokes (RANS) models \cite{alfonsi2009reynolds} and hybrid RANS-LES approaches \cite{heinz2020review}. 
Nonetheless, these approaches require a trade-off between accuracy and computational expedience. 
In addition, data assimilation algorithms are proposed to integrate the measurements with the aforementioned numerical models. 
For example, Wang et al.\cite{wang2021state} introduced an adjoint-variational data-assimilation algorithm to reconstruct the flow fields from measurements and enhance traditional methods in the sparse condition.
However, these methods also suffer from computational inefficiency, requiring iterative solutions to the Navier-Stokes equations.
To mitigate the conflicts between computational cost and simulation accuracy, it is worth investigating the end-to-end learning-based models designed for rapid inference to bridge the gap between simulations and real-world flow field environments, which have demonstrated success in diverse applications such as image processing.

In the field of image processing, the use of machine learning for super-resolution enhancement and data recovery techniques from low-resolution or sparse pixel points is developing rapidly, including algorithms based on convolutional neural networks (CNN) \cite{dong2015image, tian2020coarse, rao2023encoding}, transformer architectures \cite{chen2023activating,  yang2020learning,  luo2022bsrt}, and generative adversarial networks (GAN) \cite{you2019ct,  bell2019blind, jiang2019edge}.
While super-resolution is generally regarded as an image-based data recovery technique, these neural network frameworks have been widely applied to fluid mechanics \cite{maulik2020probabilistic, gao2021super, sun2020physics, fukami2021global, asaka2024machine, guemes2022super}, providing promising alternatives to traditional methods.

Despite the considerable achievements gained by CNN-based methodologies in enhancing the resolution of specific low-fidelity image distributions, they share a common limitation: these models are intricately designed and fine-tuned for a distinct class of under-resolved flow field data, heavily reliant on their training data characteristics.
In contrast, real-world flow field scenarios present a far more complex and unpredictable landscape. 
Sensor placement is often sparse and varied, and the data acquired is frequently contaminated with noise. 
Such conditions inevitably result in discrepancies between training datasets and real-world testing scenarios.
This limitation presents challenges in the generalization and robustness of CNN-based models, restricting their applicability in real-world environments.

Revisiting techniques from the field of image reconstruction, if we regard sparse low-resolution fluid flow data as noisy experimental measurements, then super-resolution flow field reconstruction can be considered as a natural denoising problem. 
Therefore, the diffusion model in the generation problem \cite{ratcliff2016diffusion, kingma2021variational, song2020denoising, ho2020denoising, fudge2023protein} also has the potential to reconstruct the sparse data flow field. 
In this context, Shu et al.~\cite{Shu2023PhysicsInformed} introduced a model based on the Diffusion Probabilistic Model (DDPM) \cite{ho2020denoising} for super-resolution tasks, demonstrating its versatility across various scenarios to overcome the limitations in the CNN-based model used in fluid dynamics. 

However, due to the denoising nature of DDPM-based models, rather than incorporating physics-informed constraints directly into the objective function, a physics-conditioning embedding is applied during the training phase. 
This approach allows the model to correct physics deviations in the denoised flow field iteratively during the sampling phase. 
Although effective, the reliance on the number of sampling steps to achieve minimal partial differential equation loss may inadvertently compromise the efficiency of the algorithm.
This introduces the need for fine-tuning certain hyperparameters for specific datasets to achieve accurate results, potentially straying from the original intent of developing such models. 
Moreover, despite the integration of physical condition embedding, the flow field reconstructed using DDPM occasionally fails to comply with the physical constraints on a localized scale. 

To tackle these challenges, we propose a residual-based diffusion model for flow field super-resolution in this work, inspired by Yue et al.\cite{yue2024resshift} where a diffusion model with residual shifting was presented. 
Unlike traditional DDPM-based methods, the residual-based diffusion model requires fewer sampling steps and predicts the original high-fidelity flow field directly. 
This approach enables the practical application of Physics-Informed Neural Networks (PINNs) \cite{raissi2019physics, karniadakis2021physics, cai2021physics, raissi2018hidden, li2024synthetic} by integrating the residuals of the governing equations within the objective function, leading to a brand new model called physics-informed residual diffusion (PiRD) model in this paper. 
Our model not only demonstrates robust performance in reconstructing low-fidelity data across a variety of conditions in the turbulence experiment, but also enhances both training and sampling efficiency—requiring approximately 20 epochs for training and 20 steps for sampling.

We make the following key contributions:
\textbf{(i)} We present a novel, effective and motivated approach that embeds physical information into the diffusion model to solve the problem of flow field reconstruction under sparse and noisy measurements.
\textbf{(ii)} Due to the nature of the ResShift diffusion model, our proposed method can complete the training or inference in about 20 steps, which greatly improves the calculation speed compared with \cite{Shu2023PhysicsInformed}, and it is beneficial for future application to actual real-time flow field scenarios.
\textbf{(iii)} Through some classical turbulence experiments, PiRD is proved to be more accurate and robust than the CNN-based model for flow field reconstruction on sparse data. Moreover, PiRD can handle various sensor placements and noisy measurements which are not considered in the training data.

\section{Methods}
\subsection{Problem Setup}
The primary objective of this research is to address the limitations of current CNN-based methods in reconstructing high-fidelity (HF) flow fields, denoted as \( \omega \sim U_{\text{HF}} \), from low-fidelity (LF) observations, denoted as  \( \Tilde{\omega} \sim U_{\text{LF}} \). CNN-based models like UNet, represented as \( f_{\theta} \), have shown proficiency in dealing with in-distribution low-fidelity data. 
This process can be described as:
\[
w = f_{\theta}(\Tilde{\omega}), \quad \Tilde{\omega} = \mathcal{D}(\omega)
\]
where $\mathcal{D}$ represents certain the down-sampling operation, or a certain method to degrade the high-fidelity flow field, the neural network is essentially trying to learn this operation to reserve this process, and in-distribution data means the high-fidelity data that was corrupted by this down-sampling operation $\mathcal{D}$, or in which operations that are highly similar to it. 
However, the learned model can only be applicable for the scenarios that are degraded by this specific learned  $\mathcal{D}$, as a result, if the degraded flow field is not corrupted by it, the reconstruction process still trying to reverse $\mathcal{D}$ which ultimately make the prediction inaccurate. 
Therefore, to address this problem, we build up a Markov chain between high-fidelity flow fields \(U_{\text{HF}} \) and low-fidelity flow fields \(U_{\text{LF}} \) that are independent of $\mathcal{D}$, such that we can reconstruct low-fidelity flow fields given any possible $\mathcal{D}$. 
In addition, physics-informed loss function can also be integrated into such a process to act as not only a physical constraint to make sure that the prediction follows the underlying physical law, but also a guide for the model to infer the information lost in the low-fidelity flow field and ensure robust performance across diverse and challenging scenarios.

\subsection{Physics-informed Residual Diffusion}
This section introduces our proposed model used to map the low-fidelity distribution to the high-fidelity distribution through a learned Markov chain and to ensure that the entire process follows the underlying physical law such that we can eventually produce a high-fidelity prediction that also obeys the underlying governing equation. 
In the following, we present the architecture of Physics-informed Residual Diffusion (PiRD) which is composed of two essential modules:
\textbf{(i)} Residual Shifting Diffusion Model that could build a Markov chain between LF distributions and HF distributions; 
\textbf{(ii)} Physics-informed neural networks(PINNs) for fluid mechanics that ensure the predictions obey the underlying governing equation. 

\subsubsection{Residual Shifting Diffusion Model for Vorticity Field}
Residual Shifting Diffusion Model was first proposed by Yue et al. \cite{yue2024resshift} developed for the image super-resolution task.
This method diverges from traditional diffusion approaches that transit high-fidelity images to Gaussian noise. 
Instead, it constructs a Markov chain between high-resolution images and low-resolution images by shifting the residual between them. 
We apply this method as the backbone of our task, expand this ideal and further develop it into the application of low-fidelity flow field super-resolution and sparse flow field reconstruction.
Let $\omega$ denoted as the high-fidelity flow field, ad $\Tilde{\omega}$ as its corresponding low-fidelity flow field, and let $e = \Tilde{\omega} - \omega$ denoted the residual between them, this residual does not only contains the spatially related difference between them but also contains the physical related information lost between them. 
The forward diffusion Markov chain can be described as:
\begin{align}
    q(\omega_t|w_{t-1}, \Tilde{\omega}) &= \mathcal{N} (\omega_t; \omega_{t-1} + \alpha_t e, \kappa^2\alpha_t I), \quad t = 1, 2, \ldots, T, \\
    q(\omega_t|\omega, \Tilde{\omega}) &= \mathcal{N} (\omega_t; \omega + \eta_t e, \kappa^2 \eta_t I), \quad t = 1, 2, \ldots, T,
\end{align}
where \(\{\eta_t\}_{t=1}^T\) is a shifting schedule that increases with the time step \(t\) such that a smooth transition of $\omega$ to $\Tilde{\omega}$ over a sequence of \(T\) steps can be built. 
\(\alpha_t = \eta_t - \eta_{t-1}\) for \(t > 1\) and \(\alpha_1 = \eta_1\), \(\kappa\) is the key parameter that controls the progression of noise variance. 
Therefore the reverse process estimates the posterior distribution \(p(w|\Tilde{\omega})\) using:
\begin{equation}
p(\omega|\Tilde{\omega}) = \int p(\omega_T |\Tilde{\omega}) \prod_{t=1}^T p_\theta(\omega_{t-1}|\omega_t, \Tilde{\omega}) \, dw_{1:T},
\end{equation}
Since in the forward diffusion process we have defined $\omega_T = \omega + \eta_{T} e + \kappa \sqrt{\eta_{T}} \epsilon$, and $\eta_{T}$ $\rightarrow$ 1 we can approximate $\omega_T = (\omega + e) + \kappa \epsilon = \Tilde{\omega} + \kappa \epsilon$ where $\epsilon \sim \mathcal{N}(0, I)$, consequently, we can also approximate \(p(\omega_T |\Tilde{\omega}\)) by \(\mathcal{N}(\omega_T |\Tilde{\omega}, \kappa^2 I)\). \(p_\theta(\omega_{t-1}|\omega_t, \Tilde{\omega})\) is the reverse transition kernel. 
By minimizing the negative evidence lower bound to make \(q(x_{t-1}|x_t, x_0, y_0)\) tractable, we can get below representations:
\begin{equation}
\min_\theta \sum_t D_{\text{KL}}[q(\omega_{t-1}|\omega_t, \omega, \Tilde{\omega}) \parallel p_\theta(\omega_{t-1}|\omega_t, \Tilde{\omega})],
\end{equation}
\begin{equation}
q(\omega_{t-1}|\omega_t, \omega, \Tilde{\omega}) = \mathcal{N}\left(\omega_{t-1}; \frac{\eta_{t-1}}{\eta_t} \omega_t + \frac{\alpha_t}{\eta_t} \omega, \kappa^2 \frac{\eta_{t-1}}{\eta_t} \alpha_t \mathbf{I}\right).
\end{equation}
The above reverse transition kernel shows that for each iterative backward process, we need the original high-fidelity flow field $\omega$ to infer from $\omega_{t}$ to $\omega_{t-1}$, therefore the objective function can be designed as follows:
\begin{equation}
\min_\theta \sum_t \lVert f_\theta(\omega_t, \Tilde{\omega}, t) - \omega \rVert^2_2.
\end{equation}
For a comprehensive derivation of forward diffusion process and backward inference process, and the design of the noise schedule, please see Section \hyperref[subsec:Detailed mathematical derivation of residual diffusion model]{A} of the Appendix.
\subsubsection{Physics-informed neural networks (PINNs) for fluid mechanics}
Within the realm of fluid mechanics, the application of Physics-Informed Neural Networks (PINNs) to the vorticity transport equations exemplifies a sophisticated methodology for solving parametrized partial differential equations (PDEs) that describe complex fluid behaviours. The Vorticity Transport Equation in fluid dynamics is pivotal in describing the evolution of the vorticity vector \( \omega \), which represents the curl of the fluid velocity $u$. This equation is crucial for understanding the rotational behaviors within fluid flows. In applications where comparative analysis across different flow conditions is essential, the non-dimensional form of the vorticity transport equation, incorporating the Reynolds number \( Re \), is frequently used. The Reynolds number characterizes the ratio of inertial forces to viscous forces within the fluid, providing insight into the effects of fluid viscosity across various flow regimes. The non-dimensional vorticity transport equation is expressed as:

\begin{equation}
\frac{\partial \omega}{\partial t} + (u \cdot \nabla) \omega = \frac{1}{Re} \nabla^2 \omega
\label{eq:Vorticity_Transport_Equation}
\end{equation}
In this form, the equation highlights three major phenomena: (1) The \textbf{Temporal Change} \( \frac{\partial \omega}{\partial t} \) describes the rate of change of vorticity over time. (2) The \textbf{Advection} term \( (u \cdot \nabla) \omega \) represents the transport of vorticity by the velocity field. (3) The \textbf{Diffusion} \( \frac{1}{Re} \nabla^2 \omega \) accounts for the diffusion of vorticity, moderated by the fluid's viscosity, which is inversely scaled by the Reynolds number.
The vorticity transport equation describes the dynamics of vorticity, accounting for its transport and diffusion in fluid flows. This equation sets the stage for employing Physics-Informed Neural Networks (PINNs) in fluid mechanics, leveraging these fundamental principles to enhance model predictions and simulations.
The PINNs framework for solving the vorticity transport equation integrates the physics of the problem directly into the learning process of the neural network. 
By approximating the solution, the network is trained to comply with the governing PDE across the domain $\Omega$ and throughout the time interval $[0, T]$. 
This training process is facilitated by the formulation of a composite loss function $L$, which encapsulates the adherence of the network's output to the underlying physical laws and empirical data. 
The loss function is articulated as
\begin{equation}
L = \beta_1 L_{\text{PDE}} + \beta_2 L_{\text{data}} + \beta_3 L_{\text{IC}} + \beta_4 L_{\text{BC}},
\end{equation}
where each term represents a different aspect of the learning criteria: $L_{\text{PDE}}$ penalizes deviations from the vorticity transport PDE, $L_{\text{data}}$ quantifies the discrepancies between the predictions and actual flow measurements, $L_{\text{IC}}$ and $L_{\text{BC}}$ ensure the model's output matches the initial and boundary conditions, respectively, $\beta_{i}$ denotes the weighting coefficient for each term.
To accurately model fluid behavior, PINNs incorporate initial and boundary conditions directly into the loss function.
Initial conditions set the stage for the fluid's state at the beginning of the observation period, whereas boundary conditions dictate the fluid's behaviour at the domain's edges, influencing the evolution of the flow over time. 
The precise specification of these conditions is vital for capturing the physical scenario under investigation, such as the generation and dynamics of vorticity in flows past obstacles.

An adaptive optimization algorithm, typically ADAM, is employed to minimize the loss function. 
This optimization approach adjusts the learning rate dynamically, enabling efficient convergence towards a parameter set that best represents the solution to the vorticity transport equation while respecting both the fluid dynamics and the available empirical data.

\begin{figure*}[t!]
  \centering
  \includegraphics[width=\textwidth]{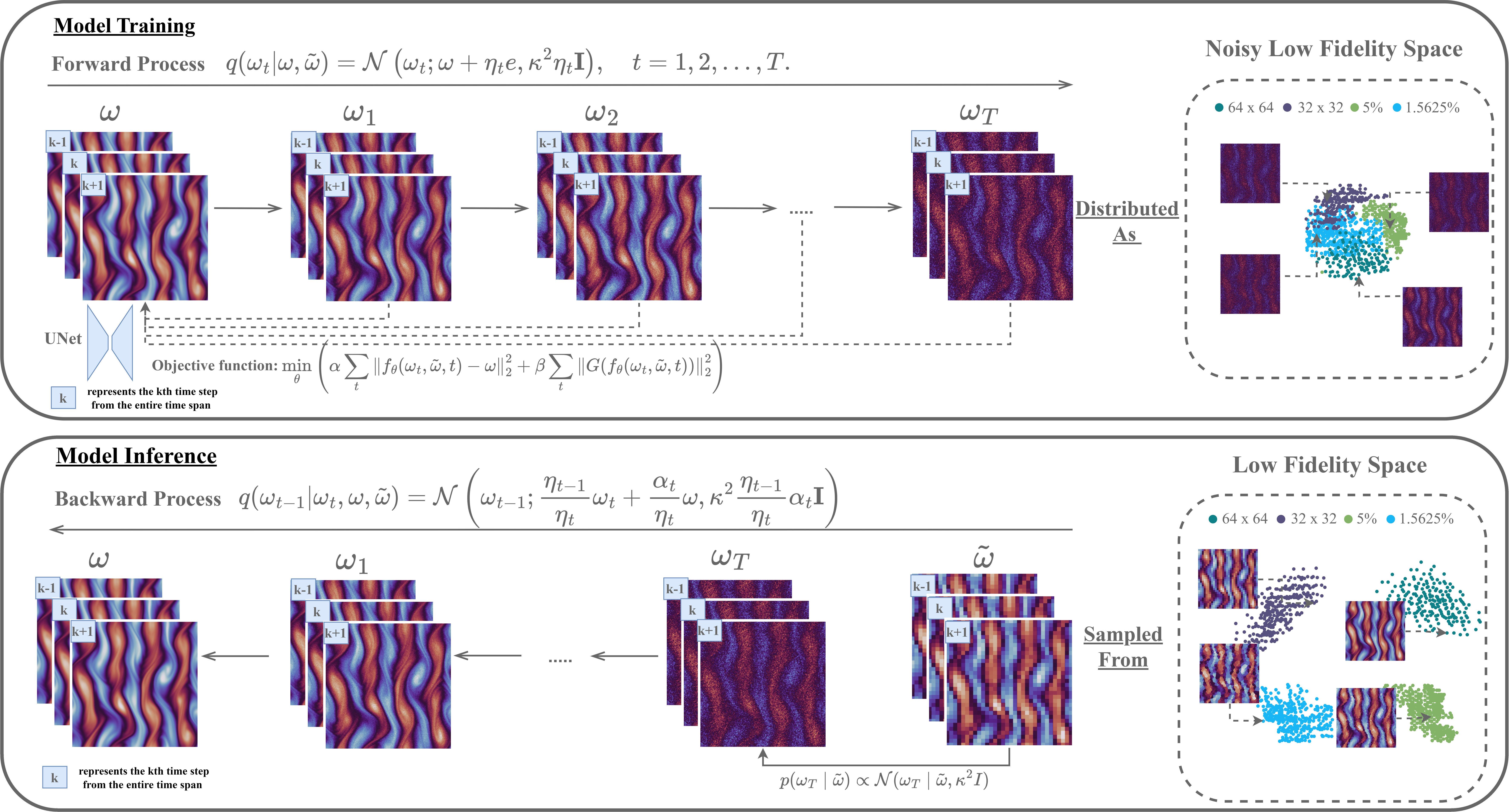} 
  \caption{The architecture of PiRD. During the forward process, a time-step $t$ is randomly drawn from $(0,T]$, then a UNet is used for predicting $\omega$ from $\omega_{t}$ with the objective function that both penalize the element-wise difference but also the physics loss. During the inference process, the LF field $\Tilde{\omega}$ is first mapped to $\omega_{T}$ then followed by the reverse Markov chain and finally predicted HF field $\omega$.}
  \label{fig:PiRD_architecture}
\end{figure*}

\subsubsection{Physics-informed Residual Diffusion}
The vanilla Residual Diffusion Model (RDM) operates by constructing a Markov chain, characterizing a residual, $e$, that quantifies the discrepancy between the high-fidelity vorticity field, $\omega$, and its low-fidelity counterpart, $\tilde{\omega}$. 
This residual is dissected into two principal components: $e_{\textit{fidelity}}$, which represents the element-wise discrepancies, and $e_{\textit{physics}}$, which captures deviations from established physical laws. 
Consequently, the relationship between $\tilde{\omega}$ and $\omega$ can be represented by the following equation:
\begin{equation}
\omega = \tilde{\omega} + e =\tilde{\omega} + (e_{\textit{fidelity}} + e_{\textit{physics}}).
\end{equation}
While $e_{\textit{fidelity}}$ plays a pivotal role in enhancing resolution, effectively addressing $e_{\textit{physics}}$ poses a significant challenge due to the complex nature of fluid dynamics, and flow field reconstruction cannot be simply seen as a mere computer vision task, which prioritizes the pixel-wise similarity. 
In many scenarios, it is possible that for a pair of predictions with the same pixel-wise mean relative error compared to the ground truth, their respective prediction has large differences in terms of physics property. 
Therefore, if the Markov chain could facilitate the inference of a high-fidelity (HF) vorticity field from a low-fidelity (LF) one while rigorously adhering to physical laws, the flow field reconstructed could make sure that it not only revert $e_{fidelity}$ but also $e_{physics}$. 
Nonetheless, realizing this ideal scenario requires two often unattainable conditions in practical applications: 
\textbf{(i)} the presence of ample training data to thoroughly represent Markov chain transitions. 
\textbf{(ii)} LF distribution representations at the termination of the chain ($X_T$) that accurately reflect all conceivable LF situations, crucial for recovering $e_{\textit{physics}}$.

To address these challenges, we introduce Physics-informed Residual Diffusion (PiRD). 
PiRD enriches the RDM framework by embedding physics-informed constraints within the model's loss function, ensuring adherence to physical laws at every step of the Markov chain.
This approach serves a dual purpose: it acts as a form of data augmentation, introducing crucial information for model learning, and it can also guarantee that each LF vorticity field is transitioned towards the HF distribution in compliance with physical principles without the concern that the physics-wise error will accumulate during the sampling process. 
Through this integration, PiRD can effectively map complex LF vorticity fields to HF fields, ensuring that subsequent transitions are physically coherent. 
As a result, the final predicted $\hat{\omega}$ not only exhibits high fidelity but also aligns with the fundamental physical principles, surmounting the inherent limitations of conventional RDM in vorticity super-resolution and reconstruction tasks.

Figure \ref{fig:PiRD_architecture} illustrates the core architecture of the proposed PiRD model. It utilizes a Markov chain methodology to facilitate the reconstruction of high-fidelity flow fields from their low-fidelity counterparts, ensuring adherence to physical laws throughout the reconstruction process.

The loss function employed in PiRD comprises two components: the element-wise loss and the PDE loss. The element-wise is the L2 norm between the prediction \( \hat{\omega} \) and the true flow field \( \omega \), normalized by the L2 norm of \( \omega \), emphasizing the accuracy of individual elements. The PDE loss ensures that the predicted field complies with the underlying governing equations, to maintain the physical plausibility. Mathematically, the loss function is expressed as:
\begin{equation}
\text{Loss} = \alpha \sum_t \|f_{\theta}(\omega_t, \tilde{\omega}, t) - \omega\|_2^2 + \beta \sum_t \|G(f_{\theta}(\omega_t, \tilde{\omega}, t))\|_2^2,
\label{eq:loss}
\end{equation}
where \( \alpha \) and \( \beta \) are the weighting parameters that balance the significance of data fidelity and physical consistency, respectively.
The differential operator \(G\), employed in the loss function, represents the dynamics captured by the non-dimensional two-dimensional (2D) vorticity transport equation for incompressible fluid flow as shown in Equation \ref{eq:Vorticity_Transport_Equation}.

In 2D incompressible flows, the velocity field \(u\) can be derived from the vorticity field \(\omega\) using the stream function \(\psi\), where \(\omega = -\nabla^2 \psi\), and \(u\) is obtained from the gradients of \(\psi\). This relationship allows the velocity field to be implicitly determined from the vorticity field, facilitating the calculation of \(G\) as:
\begin{equation}
G(f_{\theta}(\omega_t, \tilde{\omega}, t)) = \frac{\partial f_{\theta}(\omega_t, \tilde{\omega}, t)}{\partial t} + (u \cdot \nabla) f_{\theta}(\omega_t, \tilde{\omega}, t) - \frac{1}{Re} \nabla^2 f_{\theta}(\omega_t, \tilde{\omega}, t).
\end{equation}
The ideal outcome, \(G(f_{\theta}(\omega_t, \tilde{\omega}, t)) = 0\), indicates that the predicted vorticity field adheres to the dynamics governed by the vorticity transport equation. Hence, the PDE loss component is defined as the right hand side of the of Equation \ref{eq:loss}, which penalizes predictions $f_{\theta}(\omega_t, \tilde{\omega}, t)$ that deviate from its governing equation, thereby enforcing compliance with fluid dynamics principles.

To effectively compute the rate of change of vorticity over time, the model utilizes three channels of input corresponding to vorticity fields at consecutive time frames, $k-1$, $k$, and $k+1$ as shown in Figure \ref{fig:PiRD_architecture}. It's important to note that in this context, $k$ refers to the time frame within the entire dataset's time span, rather than representing a specific time step during the forward or backward process in the diffusion model. This tri-channel approach enables the approximation of $\frac{\partial \omega}{\partial k}$ using finite difference methods.

The forward process models the diffusion of the flow field from a high-fidelity state \( \omega \) to a noised low-fidelity state over discrete time steps \( t = 1, 2, \ldots, T \). This diffusion is simulated as a Gaussian process, given by:
\begin{equation}
\omega_{t} = \omega + \eta_{t}e + \kappa\sqrt{\eta_{t}}\epsilon = \omega + \eta_{t}(\omega - \Tilde{\omega}) + \kappa\sqrt{\eta_{t}}\epsilon, \quad \epsilon \sim \mathcal{N}(0, I)
\end{equation}

In the sampling process, the reverse Markov chain recovers the high-fidelity flow field starting from the low-fidelity observations \( \tilde{\omega} \). This process inverts the forward diffusion, iteratively reconstructing the previous time step \( \omega_{t-1} \) from a more noisy time step \( \omega_t \), until the original flow field \( \omega_0 \) is retrieved:

\begin{equation}
\omega_{t-1} = \frac{\eta_{t-1}}{\eta_t} \omega_t + \frac{\alpha_t}{\eta_t} f_{\theta}(\omega_t, \tilde{\omega}, t) + \kappa \sqrt{\frac{\eta_{t-1}}{\eta_t} \alpha_t} \epsilon, \quad \epsilon \sim \mathcal{N}(0, I)
\end{equation}

We choose UNet as the underlying neural network architecture because of its exceptional performance in capturing complex spatial hierarchies inherent in image-like data, which is analogous to flow fields. 
With its encoding and decoding pathways, the architecture's design is well-suited to preserving high-resolution details crucial for reconstructing the intricate structures found in vorticity fields, making it a proper choice for this task.

\section{Experimental Setup}
\begin{figure}[htbp]
  \centering
  \includegraphics[width=1\textwidth]{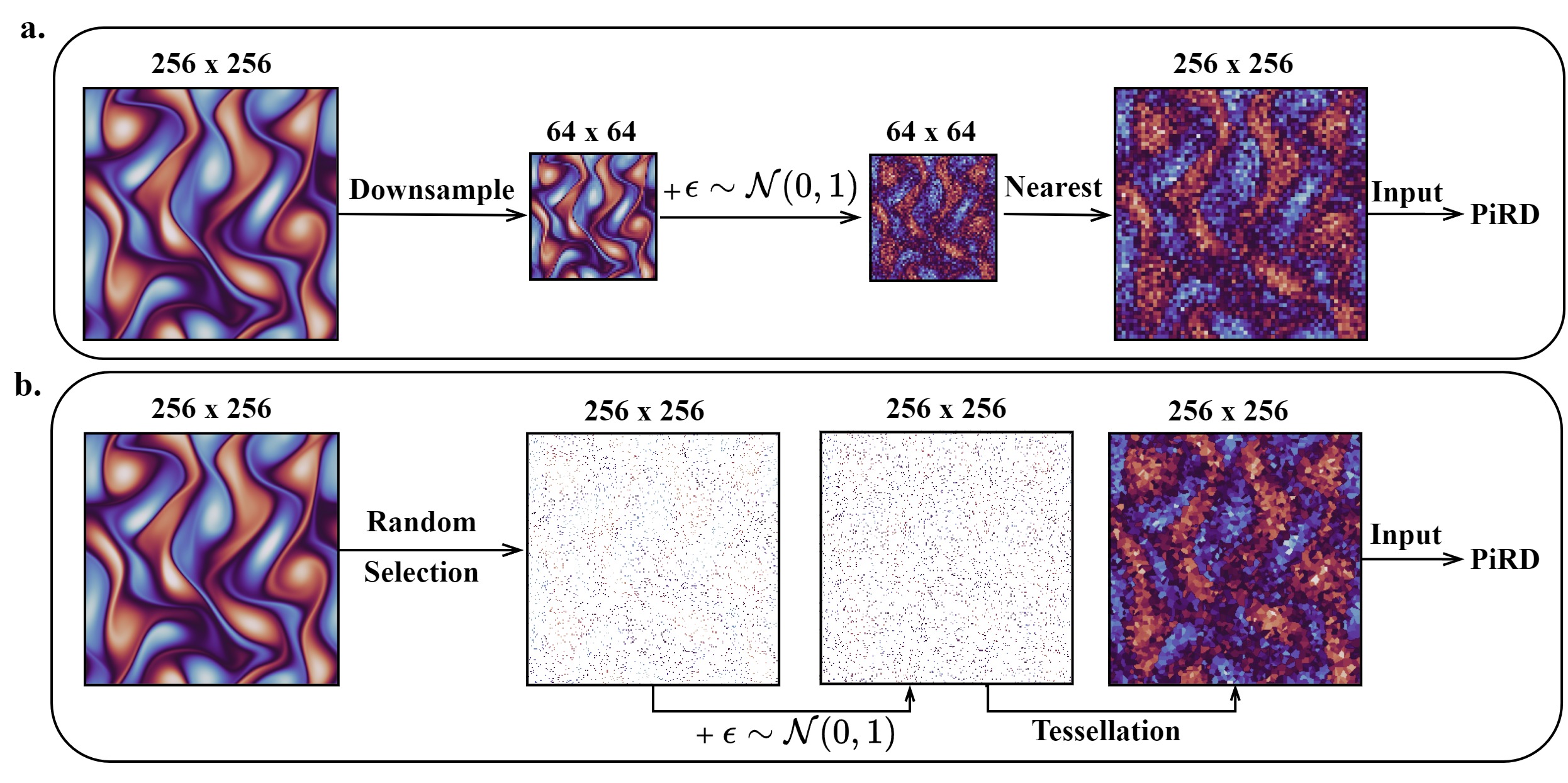}
  \captionsetup{width=1\textwidth} 
  \caption{Downsampling methods for the experiments. \textbf{(a)} illustrates an evenly down-sampling operation. The original data is initially downsampled evenly, followed by nearest interpolation to upsample it to match the desired input dimension of the model. \textbf{(b)} depicts a random selection down-sampling operation. A certain percentage of data points are randomly selected from the original dataset, after which the tesselation method is employed on the sparse dataset to achieve the desired dimension. Gaussian noise is introduced to the downsampled dataset during the noisy flow field reconstruction experiments.}
  \label{fig:experiment_design}
\end{figure}
Experiments were designed to evaluate the effectiveness and robustness of the proposed method, referred to as Physics-informed Residual Diffusion(PiRD), in reconstructing high-fidelity flow fields across various complex and practical scenarios using a two-dimensional Kolmogorov flow dataset. 
The investigation encompassed multiple tasks under different low-fidelity conditions to simulate potential challenges. 
The initial set of conditions involved simple low-fidelity scenarios where the high-fidelity flow field, originally at $256 \times 256$ pixels, was uniformly down-sampled by a scale of 4 and 8. 
This represented $4\times$ and $8\times$ up-sampling tasks, respectively.
The second scenario focused on sparse and randomly-distributed data, where only a certain percentage of the data was randomly selected for reconstruction tasks. The sparsity levels considered were $5\%$, and $1.5625\%$ of the original data, aiming to test the capability of reconstructing the high-fidelity flow field from significantly reduced data points. 
The third scenario combined low-fidelity conditions with varying levels of injected Gaussian noise. 
For this set of experiments, the low-fidelity flow field was set down-sampled by a scale of 4, and a sparse flow field with $5\%$ of the original data. 
Gaussian noise was then introduced to these low-fidelity flow fields during the testing stage at densities ranging from $20\%$ to $100\%$.
This experiment was designed to test the capacity of the PiRD in handling noisy, low-fidelity inputs. The data pre-processing methods described above can be visualized in Figure \ref{fig:experiment_design}.

\subsection{Evaluation Metrices}
For a quantitative assessment of the reconstruction, we adopt the Mean Relative Error (MRE) which is the element-wise l2 norm between the prediction and ground truth data, and divided by a normalization factor which is the l2 norm of the ground truth. 
We also calculate the PDE loss to evaluate how much our prediction deviates from the underlying physical law. These two aforesaid evaluation matrices can be described as follows: 
\begin{equation}
MRE(\hat{\omega}, \omega) = \frac{\|\hat{\omega} - \omega\|_2}{\|\omega\|_2},
\end{equation}
\begin{equation}
Loss_{\text{PDE}}(\hat{\omega}) = \frac{1}{n} \sum_{i=1}^{n} |G(\hat{\omega}_i)|^2,
\end{equation}
where $\Tilde{\omega}$ represents the predicted flow field, $\omega$ represents the original high-fidelity flow field, $G$ represents a differential operator that is applied to the predicted flow field.
\subsection{Dataset}
The dataset we consider is a 2-dimensional Kolmogorov flow same with that in Shu et al.\cite{Shu2023PhysicsInformed}, which is governed by the Navier-Stokes equations for incompressible flow. The vorticity equation that characterizes this flow is given by:
\begin{align}
    \frac{\partial \omega(x,t)}{\partial t} + u(x,t) \cdot \nabla \omega(x,t) &= \frac{1}{Re} \nabla^2 \omega(x,t) + f(x), \quad x \in (0, 2\pi)^2, \, t \in (0, T],\\
    \nabla \cdot u(x,t) &= 0, \quad x \in (0, 2\pi)^2, \, t \in (0, T],\\
    \omega(x, 0) &= \omega_0(x), \quad x \in (0, 2\pi)^2.
\end{align}
where $\omega$ represents vorticity, $u$ is the velocity field, $Re$ is the Reynolds number (set to 1000 for this dataset), and $f(x)$ is a forcing term defined as $f(x) = -4 \cos(4x^2) - 0.1\omega(x,t)$, to mitigate energy accumulation at larger scales. The boundary condition of this dataset is periodic.
The dataset contains 40 simulations, each with a temporal span of 10 seconds partitioned into 320 snapshots and a spatial resolution of $256 \times 256$ grid.

\subsection{Model Training}
The dataset was partitioned into training ($80\%$), validation ($10\%$), and testing ($10\%$) sets. 
Optimization strategies are based on a cosine annealing learning rate, the maximum number of epochs was set to 20, the maximum and minimum learning rate was set to $1e^{-3}$ and $1e^{-5}$ respectively, $T$ was set to 20, and $\kappa$ was set to 2 to ensure the generalizability of $\omega_{T}$. We set the weights for the data term and PDE term as 0.7 and $2e^{-4}$, respectively. 
The entire training process was conducted on a single RTX3090 GPU over approximately 3 hours.

\section{Results}

\subsection{Comparison with current reconstruction methods}

In this section, we conduct a quantitative analysis comparing our Physics-informed Residual Diffusion (PiRD) model against three state-of-the-art super-resolution methods: a modified UNet with ResNet block and attention block which is similar to the model proposed by Pant et al.\cite{pant2020deep}, a physics-informed Denoising Diffusion Probabilistic Model (DDPM) proposed by Shu et al. \cite{Shu2023PhysicsInformed}, and the Residual Diffusion Model (RDM) proposed by Yue et al. \cite{yue2024resshift}. 

To ensure a fair comparison, UNet, RDM, and PiRD are all trained for a reconstruction task scaling from $64 \times 64$ to $256 \times 256$. 
Notice that, unlike the DDPM-based method—which requires only the high-fidelity vorticity field for training—our PiRD model necessitates prior knowledge of the residual between high-fidelity (HR) and low-fidelity (LR) fields.

Subsequent tests evaluate the Mean Relative Error (MRE) and Partial Differential Equation (PDE) loss across four different tasks: $4\times$ up-sampling, $8\times$ up-sampling, $5\%$, and $1.5625\%$ reconstruction to $256 \times 256$ resolution. We utilize bicubic interpolation to resize the down-sampled low-fidelity dataset during training, ensuring that the input matches the desired dimensions of our output. Subsequently, for up-sampling tasks, we employ nearest interpolation, and for reconstruction tasks, we utilize Voronoi tessellation to resize the input, maintaining consistency with the desired output dimensions ($256 \times 256$). This standardized approach ensures that all testing distributions and patterns remain unseen for UNet, RDM and PiRD, facilitating a fair comparison of their performance. As the reverse process of the diffusion models are initialized with a random sample, we also iterate each batch 10 times and take the average of the flow fields as our final prediction. This approach is designed to reduce the influence of randomness in the predictive outcomes to produce a more stable and reproducible result. By computing the standard deviation over these 10 results, we can also provide the uncertainty of the predicted flow fields. For a visualization of the standard deviation of these 10 iterations and the residual of the prediction and the ground truth, we invite readers to consult the Section \hyperref[subsec:Residual]{C.5} of the Appendix. For training PiRD, $T$ was set to 20 and $\kappa$ was set to 2. For inference, we test PiRD's performance on both $T = 20$ and $T = 10$, since our experiment demonstrates that PiRD demonstrates a flexibility on $T$ due to the inherent flexibility of our variance schedule design. Results for the physics-informed DDPM are generated using the hyper-parameters recommended by the original authors, with Discrete Denoising Implicit Models (DDIM) sampling steps set at 30 and 60 respectively. The Residual Diffusion Model (RDM) without the physical loss component is trained for an ablation study, and the hyper-parameters for training the RDM are identical with those for training PiRD.

\begin{table}[t]
\centering
\begin{tabularx}{\textwidth}{@{}l*{8}{>{\centering\arraybackslash}X}@{}}
\toprule
\multirow{2}{*}{Methods} & \multicolumn{2}{c}{4x} & \multicolumn{2}{c}{8x} & \multicolumn{2}{c}{5\%} & \multicolumn{2}{c}{1.5625\%} \\
\cmidrule(lr){2-3} \cmidrule(lr){4-5} \cmidrule(lr){6-7} \cmidrule(lr){8-9}
                         & MRE & PDE & MRE & PDE & MRE & PDE & MRE & PDE \\
\midrule
PiRD - 20(ours)              & \textbf{0.1178} & \textbf{2.0663} & \textbf{0.3616} & \textbf{1.9088} & 0.2889 & \textbf{1.8443} & \underline{0.3814} & \textbf{2.1696} \\
PiRD - 10(ours)              & 0.2033 & \underline{2.1465} & 0.4726 & \underline{2.6937} & \textbf{0.1966} & \underline{1.9114} & \textbf{0.3733} & \underline{2.9808} \\
RDM - 20              & 0.1520 & 49.8802 & 0.3617 & 43.1972 & 0.3027 & 39.7595 & 0.3967 & 45.2784 \\
DDPM-based - 30         & 0.2832 & 8.0110 & 0.5251 & 7.4526 & \underline{0.2811} & 3.7741 & 0.3912 & 14.7641 \\
DDPM-based - 60         & 0.2844 & 5.1598 & 0.5258 & 4.6567 & 0.2881 & 2.1297 & 0.4378 & 3.8925 \\
UNet & \underline{0.1207} & 259.0664 & \underline{0.3995} & 699.3475 & 0.3478 & 668.0065 & 0.4981 & 998.8791 \\
\bottomrule
\end{tabularx}
\caption{Quantitative Comparison. MRE means the Mean Relative Error, PDE means Partial Differential Equation residual of the predicted vorticity field.
The best and second best results are highlighted in \textbf{bold} and \underline{underline} respectively. All the models are trained with 4x up-sampling data only. The numbers beside the methods represents the inference step (e.g. ``-10'' represents the prediction takes 10 steps to sample).}
\label{tab:model_comparison_table}
\end{table}
\begin{figure*}[t]
  \centering
  \includegraphics[width=1\textwidth]{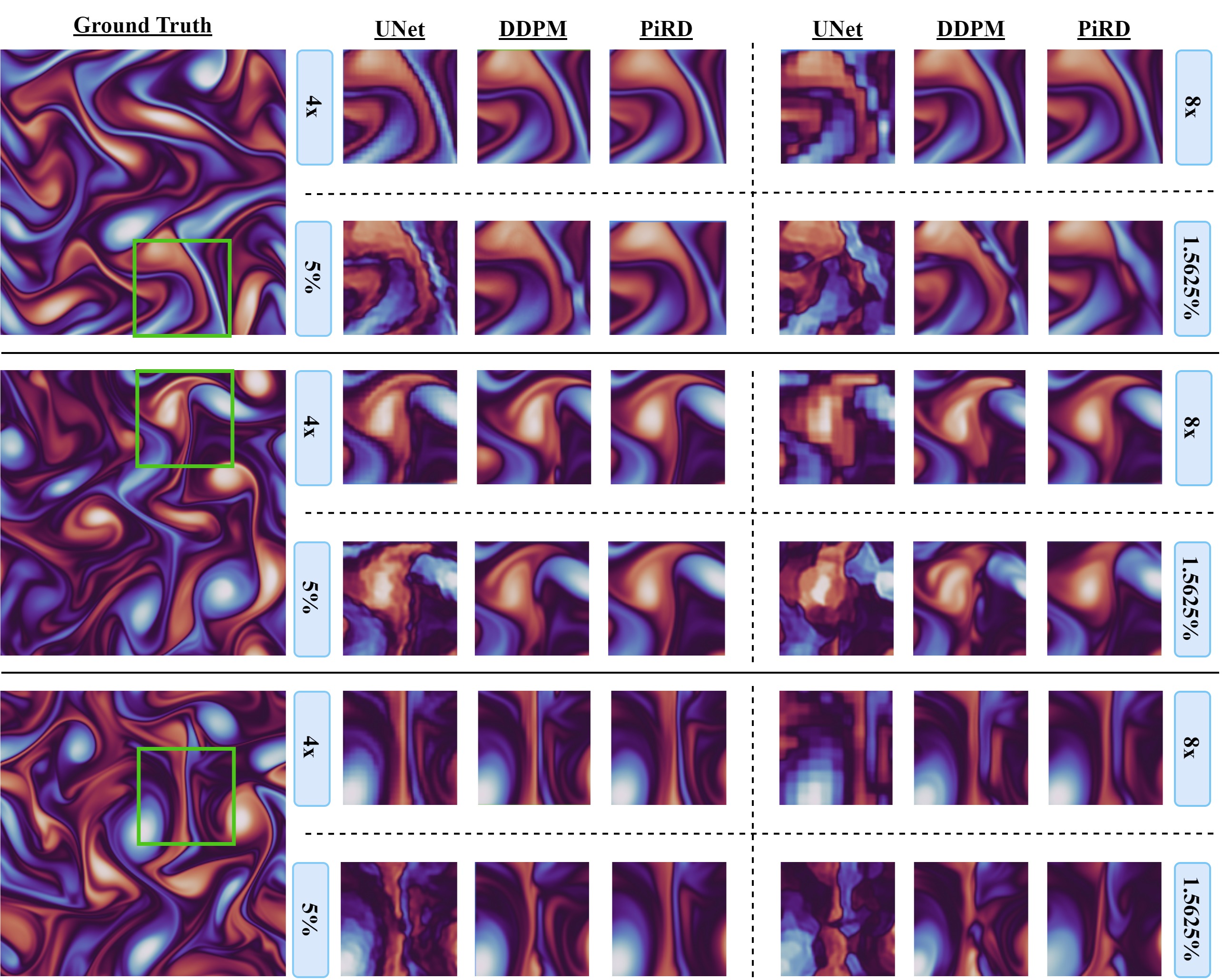}
  \caption{Flow field reconstruction from four different low-fidelity inputs (i.e., $4\times$ downsampling, $8\times$ downsampling, $5\%$ random selection, and $1.5625\%$ random selection). A zoomed-in comparison among PiRD, DDPM-based, and UNet methods for three testing cases.}
  \label{fig:image5}
\end{figure*}

As illustrated in Table \ref{tab:model_comparison_table}, UNet model struggles across all the tested scenarios particularly in terms of the PDE loss, reflecting its inability to generalize to unseen low-fidelity distributions and patterns which haven't been encountered during training. Although it achieves the second lowest MRE on the 4x up-sampling task, it performs worst on recovering the underlying physical law. The extremely high PDE loss validates our opinion that the flow field reconstruction cannot be merely seen as a computer vision task, which may result in a successful recovery of element-wise information, but unacceptable physical-wise information. This can be further validated by visualizing the zoomed-in qualitative comparison in Figure~\ref{fig:image5}. As shown, UNet model fails to predict reasonable and smooth flow fields, especially when the low-fidelity input contains randomly distributed measurements (e.g., the $5\%$ and $1.5625\%$ reconstruction tasks). 

Conversely, PiRD and DDPM-based method, consistently maintains low and reasonable PDE losses, indicating successful recovery of underlying physical laws during the reconstruction process while also minimizing mean relative errors. Notably, PiRD outperforms other methods across all scenarios in terms of both MRE and PDE loss, confirming our assumption that different low-fidelity scenarios can be collectively mapped into a noisy LF distribution, followed by the use of the learned reverse Markov chain to sample LF flow fields back to the HF distribution. Furthermore, compared to the RDM (Residual Diffusion Model), PiRD demonstrates superior performance not only in terms of PDE loss but also in achieving lower Mean Relative Error (MRE). This outcome highlights the effectiveness of incorporating the physical component into the loss function, ensuring adherence to the underlying governing equation and providing guidance for distribution learning, ultimately resulting in reduced MRE loss.

It is noteworthy that the performance of the DDPM-based method on PDE loss is dependent on the number of sampling steps employed. Specifically, increasing the steps from 30 to 60 tends to reduce PDE loss but at the expense of higher MRE. This trade-off arises because the DDPM-based method involves subtracting the PDE deviation relative to the vorticity field at each sampling step - a process that is challenging to manage without potentially compromising MRE performance.

In contrast, our PiRD model excels in predicting the high-fidelity field with both a remarkably low MRE and an exceptionally low PDE loss, demonstrating its effectiveness and robustness in the flow field reconstruction task.
For a comprehensive and varied examination of PiRD's performance, we invite readers to consult the Section \hyperref[sec:Qualitative Comparison]{C} of the Appendix, where the reconstructed flow fields with different inputs are presented.

\subsection{Evaluation on kinetic energy spectrum and vorticity distribution}

\begin{figure*}[t!]
  \centering
  \begin{subfigure}[b]{0.45\textwidth}
    \includegraphics[width=\textwidth]{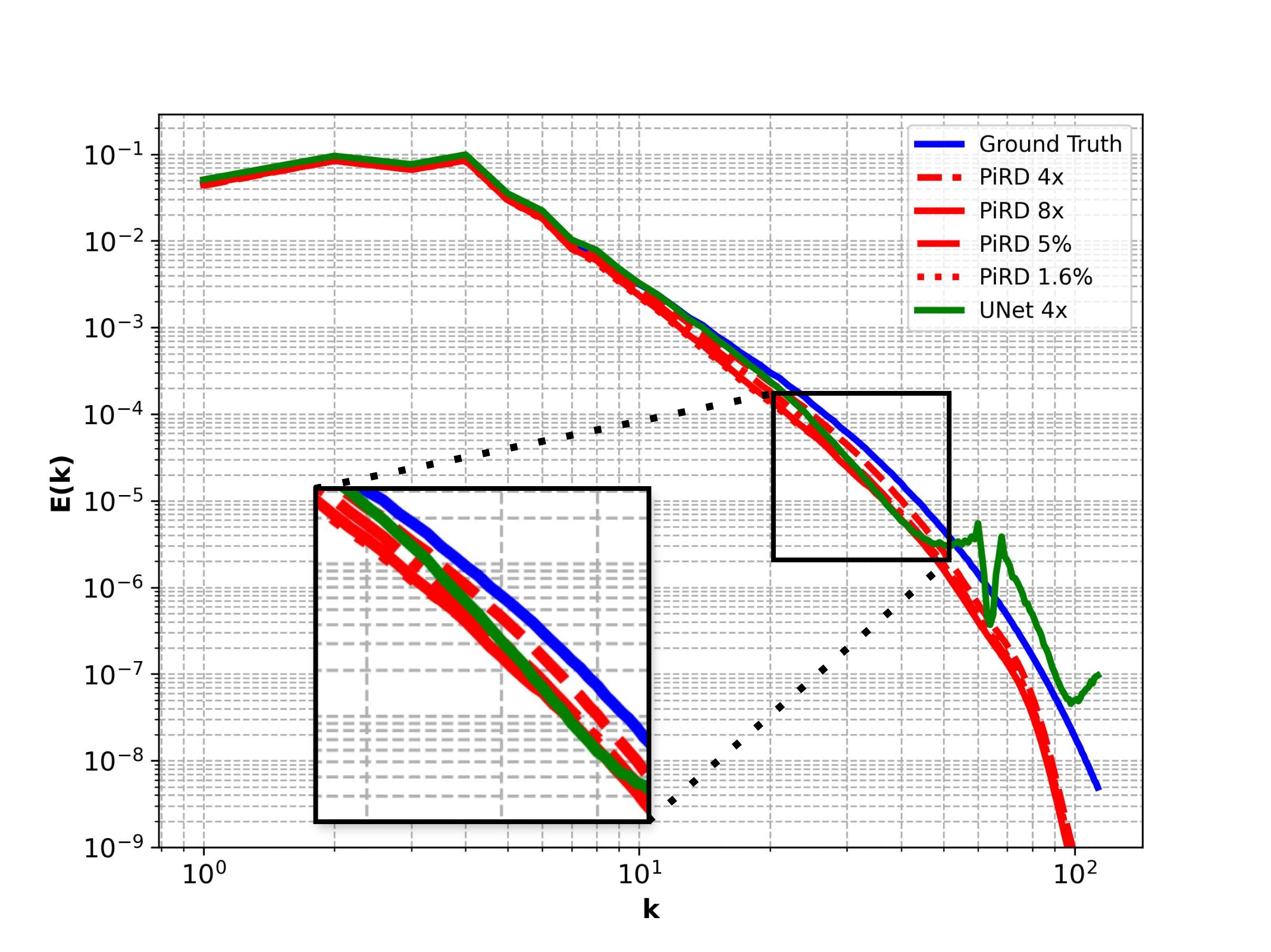}
    \caption{Kinetic energy spectrum}
    \label{fig:kinetic_spectrum}
  \end{subfigure}
  \hspace{0.2cm}
  \begin{subfigure}[b]{0.45\textwidth}
    \includegraphics[width=\textwidth]{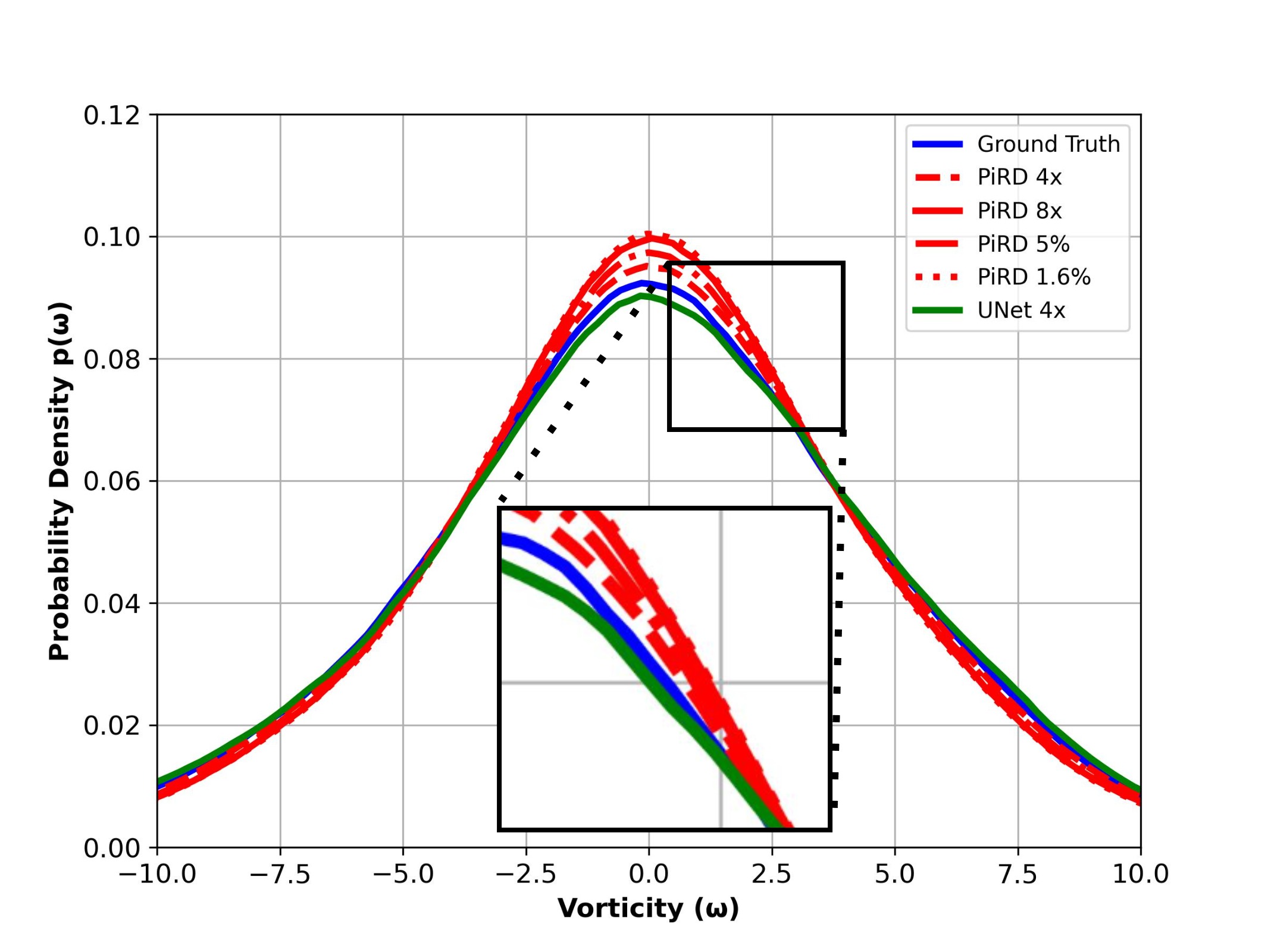}
    \caption{Vorticity distribution}
    \label{fig:vorticity_distribution}
  \end{subfigure}
  \caption{Comparison of PiRD's performance on (a) kinetic energy spectrum and (b) vorticity distribution.}
  \label{fig:both_metrics}
\end{figure*}

The kinetic energy spectrum is a crucial metric in turbulence research and is used to understand the distribution of energy among different scales within the flow field. It characterizes how kinetic energy varies with the wavenumber, revealing the flow's energy cascade from larger to smaller scales. 
It is important to highlight that despite UNet's mean relative error (MRE) being comparable to that of PiRD for the task of 4x up-sampling, as indicated in Table \ref{tab:model_comparison_table}, its performance on the kinetic energy spectrum is relatively poor, as shown in Figure \ref{fig:kinetic_spectrum}. Although it performs well in recovering large-scale vortices, it fails to retrieve the vorticity at higher wavenumber. 
This discrepancy further validates our view that the accurate recovery of element-wise data does not necessarily translate to successfully retrieving physically meaningful information.
On the contrary, our method demonstrates a robust capability to recover the energy distribution of the turbulent flow as shown in Figure \ref{fig:kinetic_spectrum}.

The spectrum obtained by our technique aligns closely with the ground truth across a range of wavenumbers, particularly in the inertial subrange, indicating that our method successfully captures the scale-dependent distribution of kinetic energy. The slight deviation at higher wavenumbers can be attributed to the inherent challenges in resolving the smallest scales of turbulence.

Furthermore, the vorticity distribution offers valuable insight into the likelihood of encountering specific vorticity magnitudes within the flow. An accurate method for capturing this distribution from coarse data suggests its ability to preserve essential characteristics of turbulence structure. These include features such as the distribution's peak and tails, which provide valuable information about underlying turbulence phenomena.

The obtained distribution as illustrates in Figure \ref{fig:vorticity_distribution}, demonstrates a strong alignment with the ground truth, particularly in accurately representing the distribution's peak and tails. These aspects are indicative of significant vorticity events within the flow.

\subsection{Evaluation on noisy low-fidelity vorticity field}

\begin{figure*}[t!]
  \centering
  \includegraphics[width=1\textwidth]{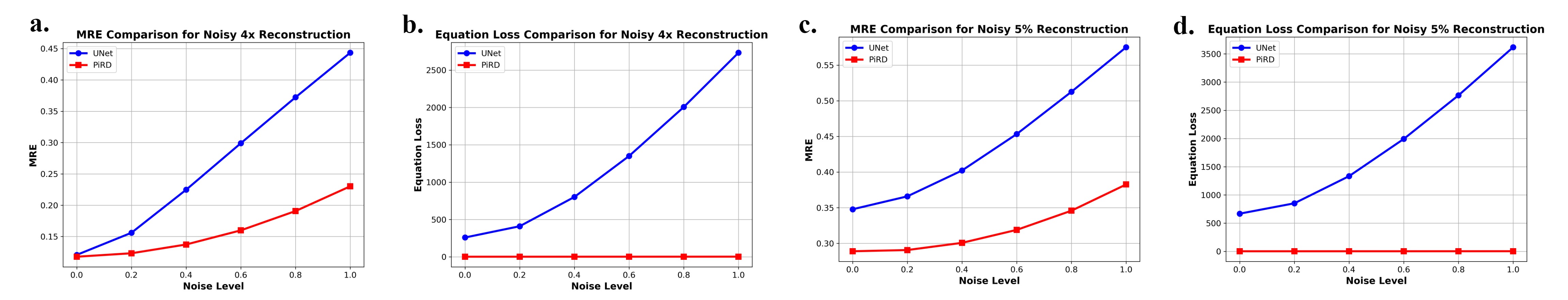}
  \caption{PiRD and UNet's performances on MRE and PDE under a progressive density of injected Gaussian noise. (\textbf{a} and \textbf{b}) illustrate the comparisons of MRE and PDE loss for PiRD and UNet on the noisy 4x reconstruction task. (\textbf{c} and \textbf{d}) illustrate the comparisons of MRE and PDE loss for PiRD and UNet on the noisy 5\% reconstruction task.}
  \label{fig:noise_MRE_PDE}
\end{figure*}

\begin{figure*}[t!]
  \centering
    \includegraphics[width=1\textwidth]{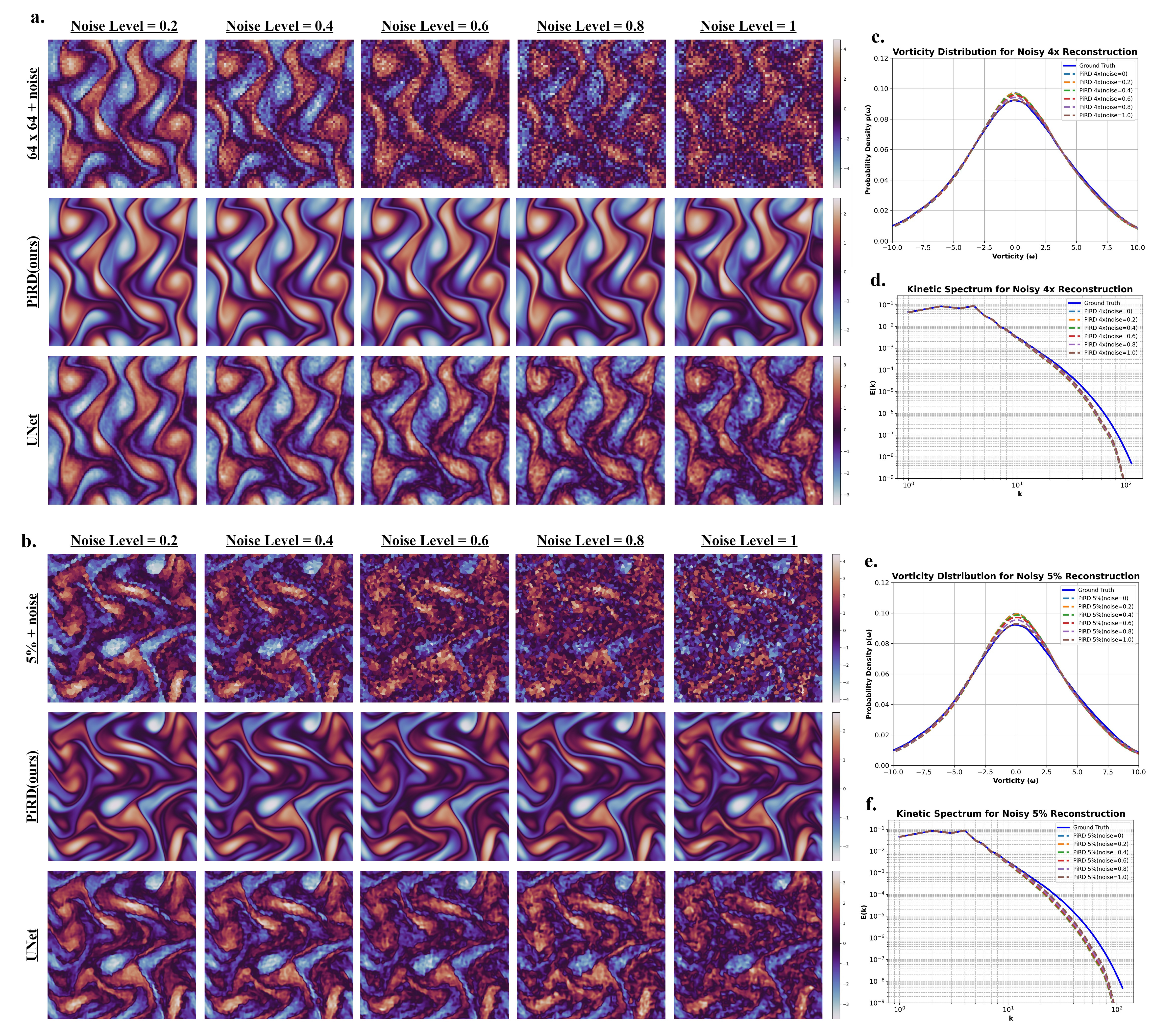}
    \label{fig:kinetic_spectrum_4}
  \caption{Reconstruction performance of PiRD and UNet under a progressive density of injected Gaussian noise. (\textbf{a} and \textbf{b}) illustrate PiRD and UNet 's performances on Noisy low-fidelity vorticity field reconstruction for 4x and 5\% senarios respectively. (\textbf{c} and \textbf{d}) illustrate PiRD's performance on vorticity distribution and kinetic energy spectrum  for 4x reconstruction task. (\textbf{e} and \textbf{f}) illustrate PiRD's performance on vorticity distribution and kinetic energy spectrum for 5\% reconstruction task. Both models are trained with clean 4x up-sampling data without any noise.}
  \label{fig:noise_spectrum}
\end{figure*}

In this section, we investigate the robustness of PiRD against a progressive density of Gaussian noise injected into the low-fidelity input as shown in Figure \ref{fig:experiment_design}, the noise injection and prediction process can be described below
\begin{equation}
\hat{\omega} = f_{\theta}(\Tilde{\omega} + \alpha \epsilon), \quad \epsilon \sim \mathcal{N}(0, I)
\end{equation}
where $\alpha$ ranges from 0.2 to 1.0 by a step size of 0.2, the low-fidelity field is down-sampled from the high-fidelity one by a scale of 4, and randomly selected 5\% of the high-fidelity data respectively, and then normalized to Gaussian distribution before the Gaussian noise is injected. 
As shown in Figure \ref{fig:noise_MRE_PDE}, as the injected noise density increases, UNet's performance degrades significantly. 
On the contrary, PiRD has stably and accurately predicted the HF field for both scenarios. 
In addition, as illustrated in Figure \ref{fig:noise_spectrum}, the kinetic energy spectrum and vorticity distribution appear identical across all noise densities compared to the case where no noise is injected. 
This observation suggests that PiRD is highly robust to noise interference. Although there is a slight degradation in MRE as the noise density increases, the performance on PDE loss, vorticity distribution, and kinetic energy spectrum remains consistent.

\section{Conclusion}

This study highlights the inherent limitations of conventional CNN-based methods in reconstructing flow fields. Specifically, these methods tend to perform optimally only on data that mirrors the distribution or down-sampling processes they were originally trained on. A mere adaptation of computer vision techniques to flow field reconstruction tasks often leads to results that, while visually plausible, may not adhere strictly to the mathematical and physical principles governing the flow dynamics.

Our proposed model, PiRD, is expected to address these challenges by establishing a Markov chain that links low-fidelity and high-fidelity flow fields. This approach not only ensures the retention of physical laws throughout the training and sampling processes but also exhibits remarkable resilience to various down-sampling operations. By prioritizing both element-wise accuracy and adherence to physical laws, PiRD may set a new benchmark in performance across experimental datasets. 
PiRD is demonstrated to be promising for flow field reconstruction tasks, merging the realms of deep learning and fluid dynamics with unprecedented efficacy.

As the first attempt to apply the Residual Diffusion Model to the flow field reconstruction problem, the proposed method still exhibits certain limitations. While PiRD demonstrates improved accuracy and robustness compared to CNN-based models when handling unseen sparse and noisy measurements, further optimization of its performance is possible when both the training and testing sets adhere to the exact same distribution. Moreover, a correlation between the standard deviation of multiple model samples and the reconstruction error can be established.

In the near future, exploring the integration of more controllable physical flow information into the diffusion model training process, as an alternative to the conventional addition of Gaussian noises, may represent a promising direction for developing a more accurate and generalizable flow generation model.

\bibliographystyle{unsrt}
\bibliography{references}
\clearpage 

\section*{Appendix}
\label{sec:Appendix}
\appendix
\renewcommand{\thesection}{A}
\renewcommand{\thesubsection}{\thesection.\arabic{subsection}}
\renewcommand{\thesubsubsection}{\thesubsection.\arabic{subsubsection}}
\section{Detailed mathematical derivation of residual diffusion model}
\label{subsec:Detailed mathematical derivation of residual diffusion model}
\subsection{Derivation of the Forward Process}
The conditional probability distribution forward process can be described as follows:
\begin{align}
q(\omega_t | \omega_{t-1}, \tilde{\omega}) &= \mathcal{N}(\omega_t; \omega_{t-1} + \alpha_t e, \kappa^2 \alpha_t I), & t &= 1, 2, \ldots, T, 
\end{align}
where its equation form can be therefore described as:
\begin{equation}
\begin{aligned}
\omega_t &= \omega_{t-1} + \alpha_{t}(\omega_t - \omega) + \kappa\sqrt{\alpha_t}\epsilon_{t} \\
&= \omega_{t-1} + \alpha_{t}e + \kappa\sqrt{\alpha_t}\epsilon_{t} \\
&= (\omega_{t-2} + \alpha_{t-1}e + \kappa\sqrt{\alpha_{t-1}}\epsilon_{t-1}) + \alpha_{t}e + \kappa\sqrt{\alpha_t}\epsilon_{t} \\
&\ldots \\
&= \omega + \sum_{i=1}^{t}\alpha_{i} e + \kappa\sum_{i=1}^{t}\sqrt{\alpha_{i}}\epsilon_{i} \\
&= \omega + \sum_{i=1}^{t}(\eta_{i} - \eta_{i-1}) e + \kappa\sum_{i=1}^{t}\sqrt{\alpha_{i}}\epsilon_{i},\\
&= \omega + \eta_{t} e + \kappa\sum_{i=1}^{t}\sqrt{\alpha_{i}}\epsilon_{i},\\
\end{aligned}
\end{equation}
where $\kappa\sum_{i=1}^{t}\sqrt{\alpha_{i}}\epsilon_{i} \sim \mathcal{N}(0, Var(\kappa\sum_{i=1}^{t}\sqrt{\alpha_{i}})I)$, according to the scaling rule of variance, we get get:
\begin{equation}
\begin{aligned}
    \kappa\sum_{i=1}^{t}\sqrt{\alpha_{i}}\epsilon_{i} &\sim \mathcal{N}(0, \kappa^2\sum_{i=1}^{t}\alpha_{i}I) \\
    &\sim \mathcal{N}(0, \kappa^2\eta_{t}I) \\
\end{aligned}    
\end{equation}
finally, we can express $\omega_{t}$ and $q(\omega_t | \omega_{t-1}, \tilde{\omega})$ as:
\begin{equation}
\begin{aligned}
    \omega_{t} &= \omega + \eta_{t} e + \kappa\sqrt{\eta_{t}}\epsilon_{i}\\
    q(\omega_t | \omega, \tilde{\omega}) &= \mathcal{N}(\omega_t; \omega + \eta_{t} e, \kappa^2 \eta_t I), & t &= 1, 2, \ldots, T, 
\end{aligned}    
\end{equation}
\subsection{Derivation of the Backward Process}
The entire backward process can be described as:
\begin{equation}
    p(\omega \mid \tilde{\omega}) = \int p(\omega_{T} \mid \tilde{\omega}) \prod_{t=1}^{T} p_{\theta}(\omega_{t-1} \mid \omega_{t}, \tilde{\omega}) \, d\omega_{1:T}
\end{equation}
where $ p(\omega_{T} \mid \tilde{\omega}) \propto \mathcal{N}(\omega_T \mid \Tilde{\omega}, \kappa^2I)$,The optimization for $\theta$ is achieved by minimizing the negative evidence lower bound:
\begin{equation}
\text{min}_{\theta}\sum_{t}D_{KL}[q(\omega_{t-1} \mid \omega_{t}, \omega, \Tilde{\omega})||p_{\theta}(\omega_{t-1} \mid \omega_{t}, \Tilde{\omega})]    
\end{equation}
The according to Bayes's rule, $q(\omega_{t-1} \mid \omega_{t}, \omega, \Tilde{\omega}) \propto q(\omega_{t} \mid \omega_{t-1}, \Tilde{\omega}) q(\omega_{t-1} \mid \omega, \Tilde{\omega})$, where
\begin{equation}
\begin{aligned}
    q(\omega_{t} \mid \omega_{t-1}, \Tilde{\omega}) &= \mathcal{N}(\omega_{t};\omega_{t-1} + \alpha_{t}e_{0}, \kappa^2 \alpha_{t} I )\\
    q(\omega_{t-1} \mid \omega, \Tilde{\omega}) &= \mathcal{N}(\omega_{t-1}; \omega + \eta_{t-1}e_{0}, \kappa^2 \eta_{t-1} I)
\end{aligned}
\end{equation}
Then extract the exponent component of the quadratic form of $q(\omega_{t-1} \mid \omega_{t}, \omega, \Tilde{\omega})$, we can get:\\
\begin{equation}
\begin{aligned}
    &-\frac{1}{2}\left[\frac{1}{\kappa^2 \alpha_{t}} + \frac{1}{\kappa^2 \eta_{t-1}}\right]\omega_{t-1} \omega_{t-1}^T + \left[ \frac{\omega_{t} - \alpha_{t}e_{0}}{\kappa^2 \alpha_{t}} + \frac{\omega + \eta_{t-1}e_{0}}{\kappa^2\eta_{t-1}}\right]\omega_{t-1}^T + C \\
    &= - \frac{(\omega_{t-1}-\mu)(\omega_{t-1}-\mu)^T}{2\lambda^2} + C
\end{aligned}
\end{equation}
where $\mu = \frac{\eta_{t-1}}{\eta{t}} \omega_{t} + \frac{\alpha_{t}}{\eta_{t}} \omega$ and $\lambda^2 = \kappa^2 \frac{\eta_{t-1}}{\eta_{t}} \alpha_{t}$, $C$ is the constant.

\subsection{Noise Schedule Design}
\label{subsec:Noise Schedule}
Recall that  \(\{\eta_t\}_{t=1}^T\) is the shifting schedule that increases with the time step \(t\) such that a smooth transition of $\omega$ to $\Tilde{\omega}$ over a sequence of \(T\) steps can be built. The $\eta_1$ should satisfies that $\eta_1 \rightarrow 0$, therefore $\eta_1$ is set to be the minimum value between $(0.04/\kappa)^2$ and 0.001. The $\eta_T$ should satisfies that $\eta_T \rightarrow 1$, therefore $\eta_T$ is set equal to 0.999. The intermediate time step $\eta_{t}$ is designed as follows:
\begin{equation}
\begin{aligned}
    \sqrt{\eta_t} = \sqrt{\eta_1} \times b_{0}^{\beta_{t}}, \quad t = 2, \dots,T-1
\end{aligned}
\end{equation}
where
\begin{equation}
\begin{aligned}
    \beta_{t} = (\frac{t-1}{T-1})^p \times (T-1), b_{0} = exp\left[\frac{1}{2(T-1)} log \frac{\eta_T}{\eta_1} \right].
\end{aligned}
\end{equation}
$p$ is fixed at 0.3 for our experiments.
\appendix
\renewcommand{\thesection}{B}
\renewcommand{\thesubsection}{\thesection.\arabic{subsection}}
\renewcommand{\thesubsubsection}{\thesubsection.\arabic{subsubsection}}
\section{Ablation Studies}
\subsection{PiRD's performance under different training configurations}
\label{subsec:Configs}
\begin{table}[htbp]
\centering
\begin{tabularx}{\textwidth}{@{}lcc *{2}{X} *{2}{X} *{2}{X} *{2}{X}@{}} 
\toprule
\multicolumn{3}{c}{Configurations} & \multicolumn{2}{c}{4x} & \multicolumn{2}{c}{8x} & \multicolumn{2}{c}{5\%} & \multicolumn{2}{c}{1.5625\%} \\
\cmidrule(lr){1-3} \cmidrule(lr){4-5} \cmidrule(lr){6-7} \cmidrule(lr){8-9} \cmidrule(lr){10-11}
& T & kappa & MRE & PDE & MRE & PDE & MRE & PDE & MRE & PDE \\
\midrule
& 10            & 2         & 0.1198 & 1.8086 & 0.3619 & 1.8025 & 0.2885 & 1.6290 & 0.3856 & 1.9341 \\
& 15            & 2        & 0.1196 & 1.9842 & 0.3601 & 1.8103 & 0.2910 & 1.8470 & 0.3866 & 2.0792 \\
& 20            & 2         & 0.1186 & 2.0727 & 0.3616 & 1.9088 & 0.2889 & 1.8443 & 0.3814 & 2.1696 \\
& 30            & 2         & 0.1224 & 1.8243 & 0.3590 & 1.7667 & 0.2922 & 1.7010 & 0.3866 & 2.0133 \\
\midrule
\midrule
&               & 1        & 0.0847 & 1.9991  & 0.3648 & 1.9357 & 0.2869 & 1.8438 & 0.4045 & 2.8366 \\
& 20            & 3        & 0.1539 & 2.1359 & 0.3627 & 2.0230 & 0.2980 & 2.0774 & 0.3776 & 2.1410 \\
&               & 4        & 0.2011 & 3.3922 & 0.3795 & 2.8995 & 0.3054 & 3.4197 & 0.3767 & 3.2438 \\
\bottomrule
\end{tabularx}
\label{tab:all_configurations}
\caption{PiRD's performance under different training configurations}
\end{table}

\subsection{PiRD's performance under different noise level}
\label{subsec:Noise Levels}
\begin{table}[htbp]
\centering
\begin{tabularx}{\textwidth}{@{}l*{8}{X}@{}}
\toprule
\multirow{3}{*}{Noise Level} & \multicolumn{4}{c}{MRE}                                    & \multicolumn{4}{c}{Equation Loss}                            \\ 
\cmidrule(lr){2-5} \cmidrule(l){6-9} 
                             & \multicolumn{2}{c}{4x}             & \multicolumn{2}{c}{$5\%$}      & \multicolumn{2}{c}{4x}             & \multicolumn{2}{c}{$5\%$}      \\ 
\cmidrule(r){2-3} \cmidrule(lr){4-5} \cmidrule(r){6-7} \cmidrule(l){8-9} 
                             & UNet           & PiRD           & UNet          & PiRD          & UNet           & PiRD           & UNet          & PiRD          \\ 
\midrule
0.0                          & 0.1207         & 0.1178         & 0.3478        & 0.2889        & 259.0664         & 2.0663         & 668.0065      & 1.8443
\\
0.2                          & 0.1560         & 0.1233        & 0.3659        & 0.2906        & 409.8747         & 2.0963         & 851.9433     & 1.8587        \\
0.4                          & 0.2247         & 0.1373         & 0.4022        & 0.3007        & 801.6868         & 2.1556         &  1332.9699     & 1.9334        \\
0.6                          & 0.2990         & 0.1599           & 0.4533        & 0.3189        & 1350.3967        & 2.2641         & 1992.5789     & 2.0817        \\
0.8                          & 0.3723         & 0.1907         & 0.5126        & 0.3458        & 2006.1515        & 2.4555         & 2765.8066     & 2.3558        \\
1.0                          & 0.4435         & 0.2302         &  0.5751        & 0.3826        & 2733.3027        & 2.7867         & 3618.42871     & 2.8328        \\
\bottomrule
\end{tabularx}
\caption{Comparison of UNet and PiRD performance for 4x and $5\%$ reconstruction tasks at various noise levels. The models are trained with clean 4x up-sampling data without any noise. }
\label{tab:noise_comparison}
\end{table}
\appendix
\renewcommand{\thesection}{C}
\renewcommand{\thesubsection}{\thesection.\arabic{subsection}}
\renewcommand{\thesubsubsection}{\thesubsection.\arabic{subsubsection}}
\section{Qualitative comparisons}
\label{sec:Qualitative Comparison}
\subsection{Qualitative comparisons among methods for 4x upsampling task}
\begin{figure*}[htbp] 
  \centering
  \includegraphics[width=1\textwidth]{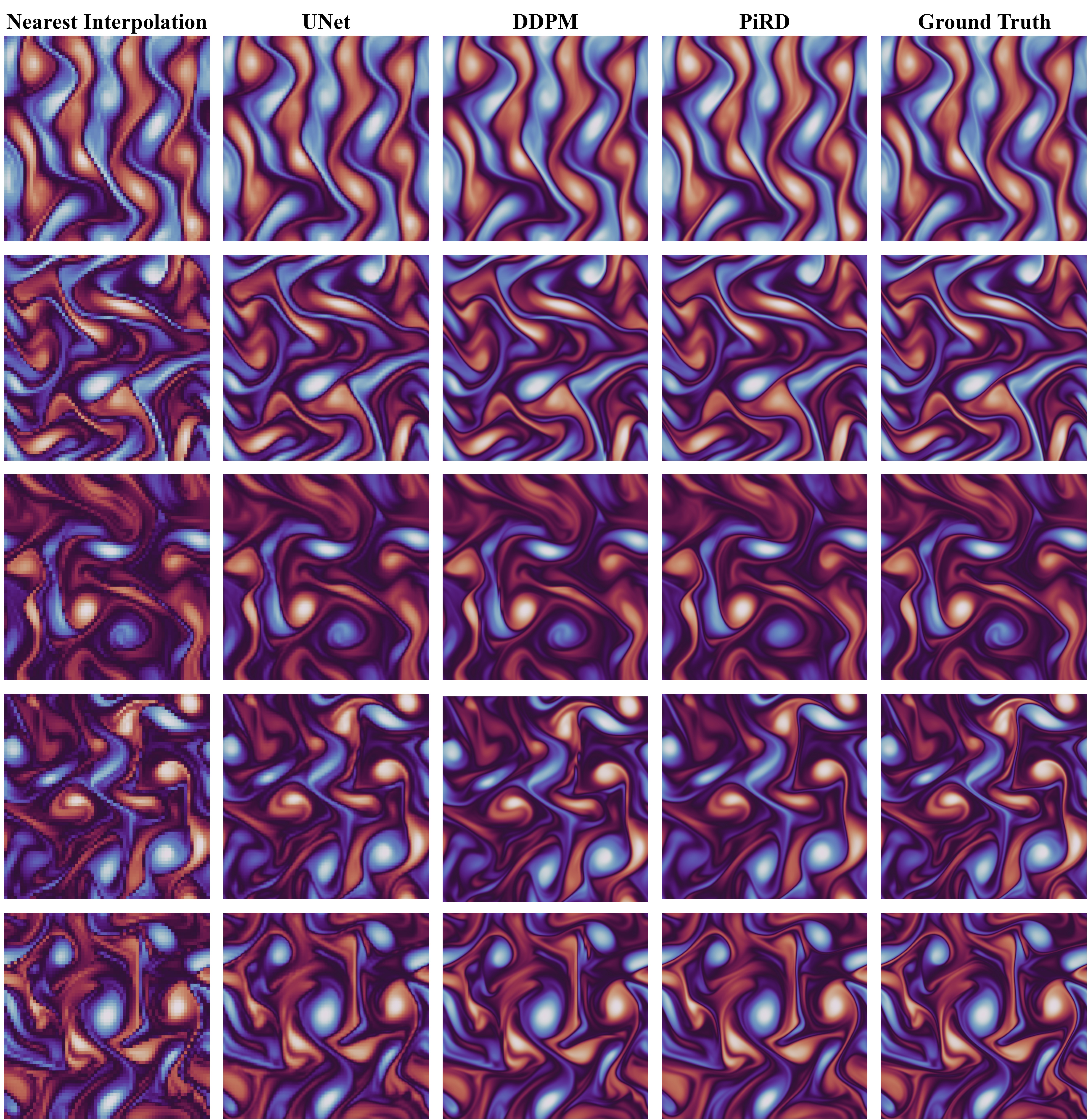}
  \captionsetup{width=1\textwidth} 
  \caption{\text{PiRD's performance on 4x upsampling task compares with other benchmarks.}}
  \label{fig:show4}
\end{figure*}
\clearpage 
\subsection{Qualitative comparisons among methods for 8x upsampling task}
\begin{figure*}[h]
  \centering
  \includegraphics[width=1\textwidth]{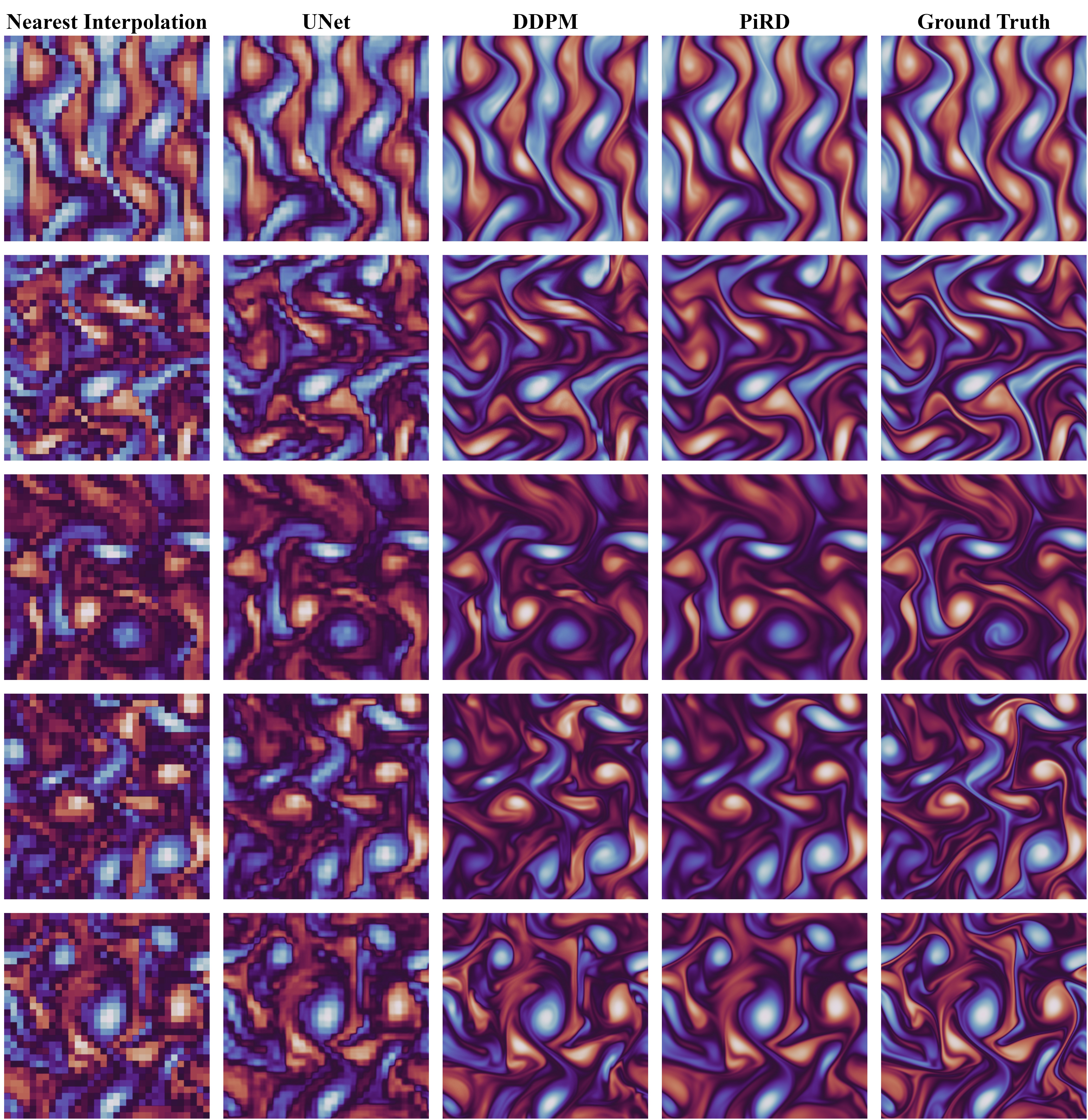}
  \captionsetup{width=1\textwidth} 
  \caption{\text{PiRD's performance on 8x upsampling task compares with other benchmarks.}}
  \label{fig:show8}
\end{figure*}
\clearpage
\subsection{Qualitative comparisons among methods for 5\% reconstruction task}
\begin{figure*}[h]
  \centering
  \includegraphics[width=1\textwidth]{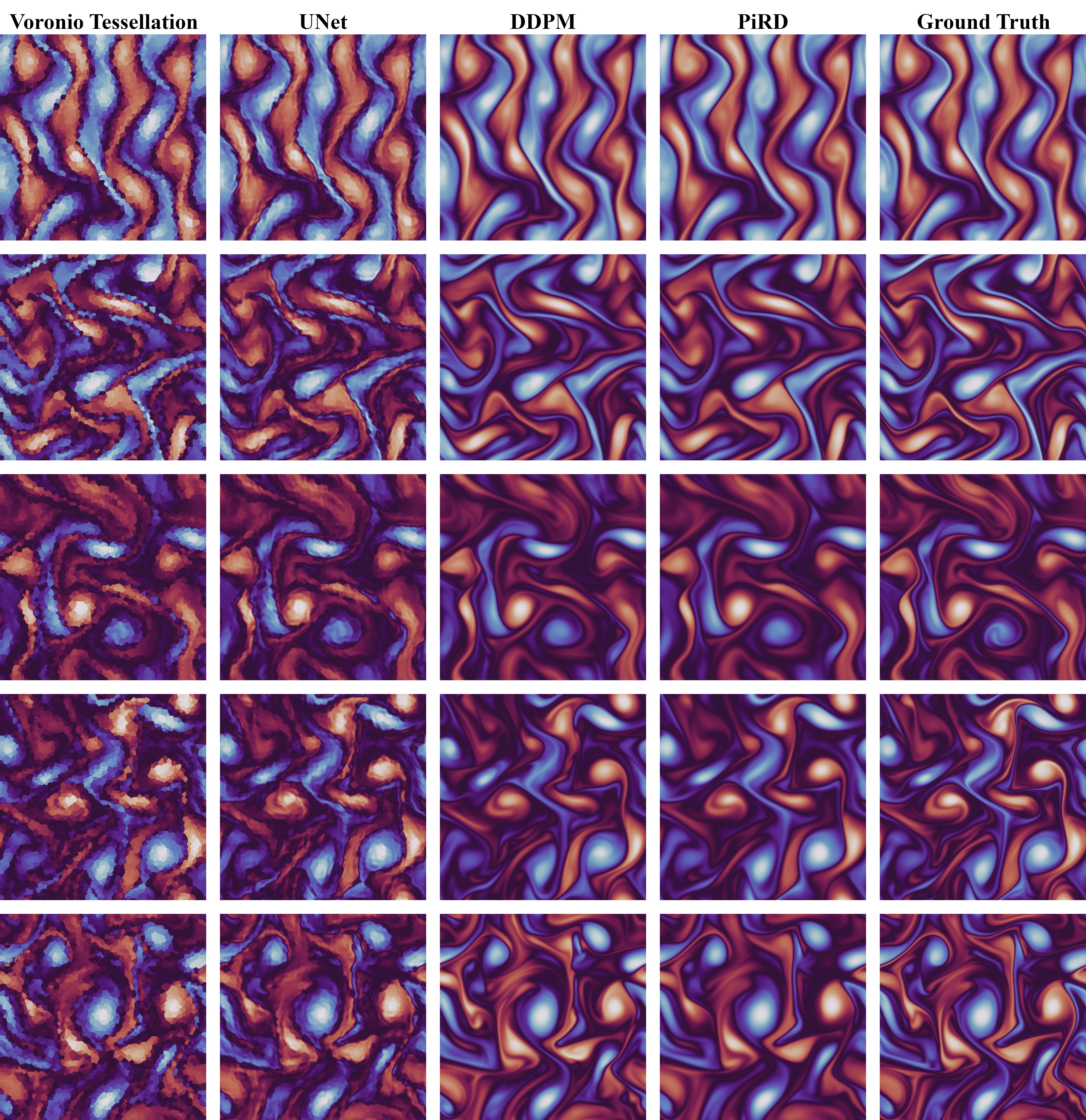}
  \captionsetup{width=1\textwidth} 
  \caption{PiRD's performance on 5\% reconstruction task compares with other benchmarks.}
  \label{fig:show0.05}
\end{figure*}
\clearpage
\subsection{Qualitative comparisons among methods for 1.5625\% reconstruction task}
\begin{figure*}[h]
  \centering
  \includegraphics[width=1\textwidth]{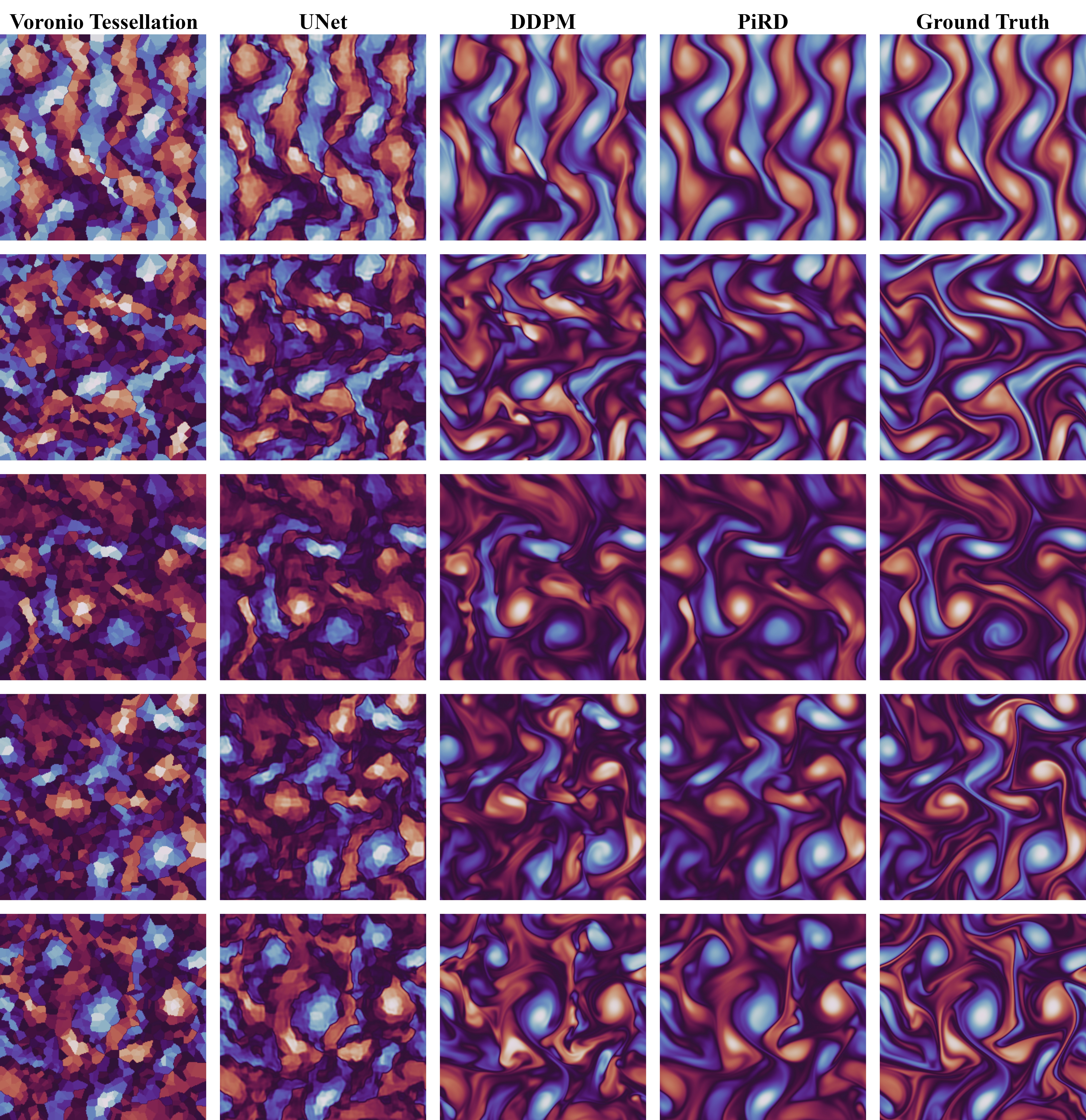}
  \captionsetup{width=1\textwidth} 
  \caption{PiRD's performance on 1.5625\% reconstruction task compares with other benchmarks.}
  \label{fig:show0.015625}
\end{figure*}
\clearpage 

\subsection{Residual vs standard deviation of PiRD's 10 samples}
\label{subsec:Residual}
\begin{figure*}[h]
  \centering
  \includegraphics[width=1\textwidth]{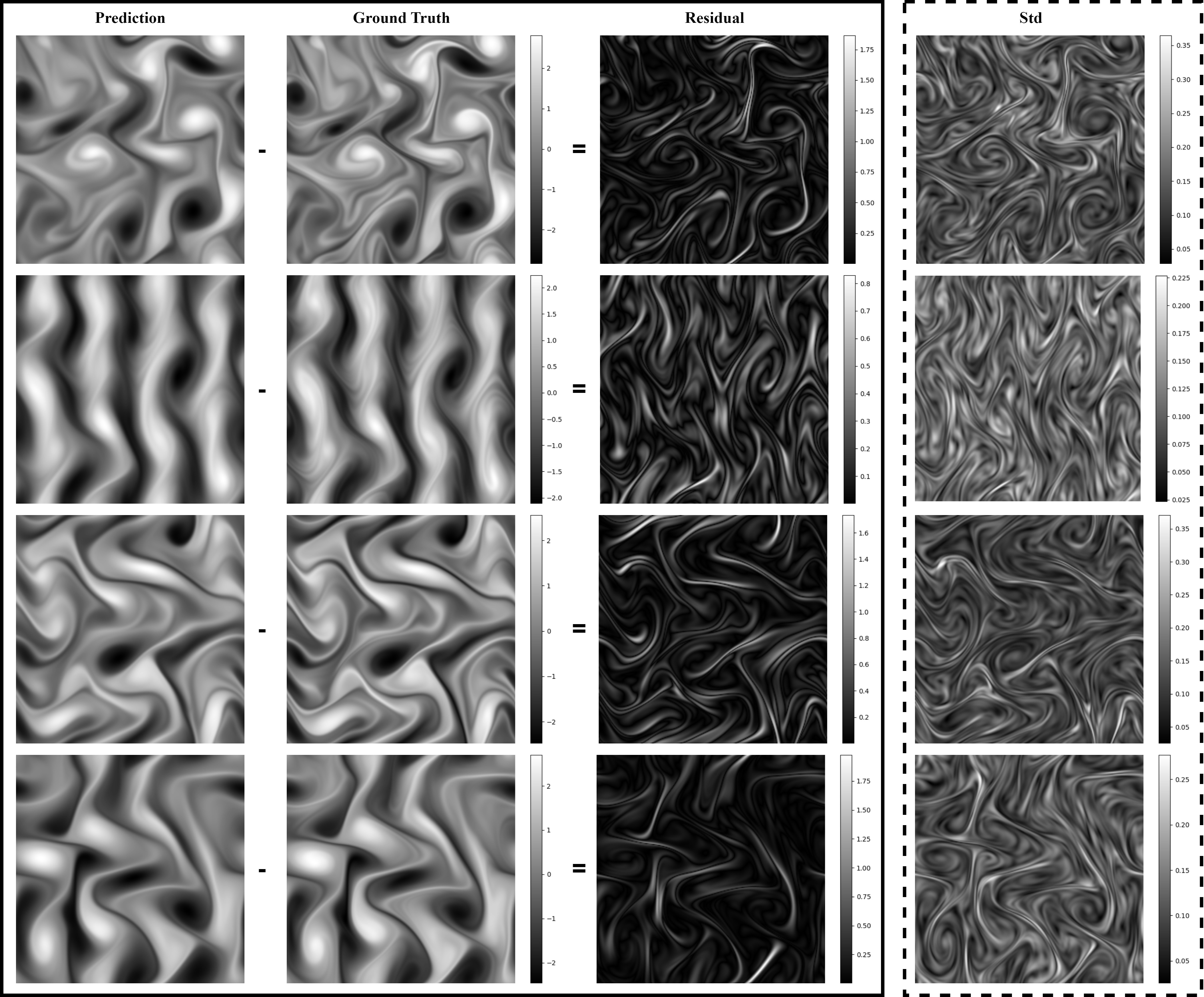}
  \captionsetup{width=1\textwidth} 
  \caption{Comparison between the standard deviation of PiRD's 10 samples and the residual between PiRD's prediction and ground truth.}
  \label{fig:residual}
\end{figure*}
\clearpage 

\end{document}